\documentclass{article}
\font\cero=cmss10 scaled 1728 
\usepackage{mathtools}
\DeclarePairedDelimiter\ket{\lvert}{\rangle}
\DeclarePairedDelimiterX\braket[2]{\langle}{\rangle}{#1 \delimsize\vert #2}
\usepackage{setspace}
\usepackage{float}
\usepackage{physics}
\usepackage{tabularx} 
\usepackage{tensor}
\usepackage{verbatim}
\usepackage{multirow}
\usepackage{amsfonts}
\usepackage{graphicx}
\usepackage{amssymb}
\usepackage{amsmath}
\usepackage{float}
\usepackage{subfig}
\usepackage{caption}
\usepackage{comment}
\usepackage{mathtools}
\DeclarePairedDelimiter\ceil{\lceil}{\rceil}

\newcommand*\diff{\mathop{}\!\mathrm{d}}
\newcommand*\Diff[1]{\mathop{}\!\mathrm{d^#1}}
\excludecomment{toexclude}
\begin{document}
\begin{flushleft}
{\cero 
The propagator field theory revisited: a Lorentz symmetry breaking approach
}\\
\end{flushleft} 
{\sf R. Cartas-Fuentevilla$^{1}$}\\
{\sf S. Gonz\'alez-Salud$^{2}$, R. B\'arcena-Ramos$^{2}$}\\
{\sf J. Berra-Montiel$^{3}$}\\
{\it 1. Instituto de F\'{\i}sica, Universidad Aut\'onoma de Puebla,
Apartado postal J-48 72570 Puebla Pue., M\'exico.} \\
{\it 2. Facultad del Ciencias F\'isico-Matem\'aticas, Benem\'erita Universidad Aut\'onoma de Puebla,
Apartado postal 1152 72570 Puebla Pue., M\'exico.} \\
{\it 3 Facultad de Ciencias, Universidad Aut\'onoma de San Luis Potos\'{\i} Campus Pedregal, Av. Parque Chapultepec 1610, Col. Privadas del Pedregal,
78217 San Luis Potos\'{\i}, SLP, M\'exico; and,  Dipartimento di Fisica ``Ettore Pancini", Universit\'a degli studi di Napoli ``Federico II", Complesso Univ. Monte S. Angelo, I-80126 Napoli, Italy.}\\

ABSTRACT: 

It is well known that the propagator for a massive scalar field is ill-defined in the coordinate space for $d\geq2$, in particular it diverges at the light-cone; we show that by using Lorentz symmetry breaking weighted measures, an infinite family of propagators can be constructed in an in\-finite\-simal strip near the light-cone, which are labeled by the weight of the measure; hence, the results will provide a finite quantum amplitude for a massive particle for propagating on the light-cone.
The propagators regarded as smooth two-points functions, increase within a region smaller than the Compton wavelength, and decrease beyond that wavelength, and eventually drop off for large arguments.
Although the time ordered propagators retain negative values regions for arbitrary values of the weight $s$ for the measures, the restriction $2<s\leq d+1$ will guarantee the positivity  for the amplitudes near the light cone.
 \\


\section{Antecedents, motivations, and results}

Recently we have initiated a program for constructing finite quantum field theories (QFT) by using Lorentz symmetry breaking weighted measures as regulators of divergences that plague the standard formulations \cite{ugalde}; the results, although incipient, allow us to glimpse that the program is promissory. The main results obtained in that reference are, first, finite vacuum expectation va\-lues for the energy without invoking normal ordering of operators, and second, finite smooth distributions for the field commutators that remain finite even at short distances. In this letter we develop further our program by reconsidering the propagator theory for a massive field, from the perspective and criticism of our scheme; the results are in order: we are able of constructing an infinite family of propagators, labeled by the weight of the measure, that are finite and well-defined in a region surrounding the light-cone,  including the light-cone itself. As a result, there is a finite amplitude for a massive field for propagating on the light-cone, in contrast to the standard description, which involves the usual light-cone divergences. Since the amplitudes are automatically finite, no subtractions are needed.

Two dimensional QFT's have a long history, with applications in string theory, statistical mechanics, particle physics, etc; they work as a laboratory for testing ideas in both, perturbative and non-perturbative regimes, and then to extend the study to higher dimensions (see for example \cite{elcio}). With these motivations, 
the case for $1+1$ QFT is developed in detail, yielding analytical expressions for the propagators in the coordinate space as functions of the mass and the weight of the measure with broken Lorentz symmetry. Preliminary results in higher dimensions are outlined.

The profile of the propagators, including their critical points, domains of increase and decrease, and regions where the amplitudes take on negative values, can be understood as a competition between the separation of the two correlated points in the quantum propagation and the Compton wavelength; moreover, these amplitudes fully satisfy the principle of microcausality. By imposing time ordering in the construction of the Feynman propagators, regions with dominant negative amplitudes inside, on, and outside the light-cone are effectively eliminated. Although regions with negative amplitudes persist for the Feynman propagators for certain values of the weight for the measures with broken Lorentz symmetry, these amplitudes may be considered admissible as small fluctuations. However, our scheme allows us to select a particular weighted measure for which all amplitudes remain positive definite; such values of the parameter $s$ lie within an interval defined for the background dimension.

In the scheme at hand the violation of the relativistic invariance comes from a deformation of the standard relativistic dispersion relation, namely, $(w_k^2)^s=(\vec{p}\:^2+m^2)^s$, where by simplicity we consider that $s$ is a real parameter; this simple deformation has a deep effect in the search of finite QFT's. The term $w_k^{2s}$ denotes the weighted measure in the definition of quantum fields, leading to a bifurcation between standard QFT with relativistic invariance, which exhibits a proliferation of divergences for $s=1$, and (an infinite family of) deformed QFT's with divergences controlled for $s\neq1$.

Modified dispersion relations allow us to incorporate effects beyond the standard QFT; for exam\-ple, the anomalies in the beta-decay spectrum and in the cosmic rays physics, can be associated with violations of the Lorentz symmetry at certain energy scale \cite{carmona}. Similarly the formulation of the doubly special relativity requires the deformation of the standard dispersion relation in order to incorporate a minimum length, namely, the Planck length; such a deformation induces a nonlinear modification of the Lorentz symmetry \cite{smolin}. Furthermore, it is well known that quantum gravity fluctuations can generate deformations in dispersion relations, thereby modifying the dynamics of particles \cite{gambini, amelino, morales, ellis, coleman}.

Deformation parameters for the measure have been previously discussed in the literature as a\-na\-lytical parameters in quantum field theory. Specifically, through a regularization process involving the exponen\-tia\-tion of the measure and the analytic continuation of the Riemann zeta function and Epstein's generalized zeta function, one can reproduce the well-known result for the  Casimir force between two conducting metallic plates \cite{ruggiero}. However, this regularization scheme is not established as a Lorentz symmetry breaking scheme like the approach being discussed. Instead it acts as an analytical regulator that does not require subtractions. In this regard, the current scheme avoids the need for subtractions, since the amplitudes are inherently finite.
Note that our real parameter $s$ can be generalized also to an analytical parameter $s_{1}+i s_{2}$, but this case will be explored elsewhere.


\section{Weighted measures with broken Lorentz symmetry}

We study for simplicity a complex scalar field in $D+1$ dimensions with Lagrangian
${\cal L}= \partial_{\mu}\varphi\partial^{\mu}\overline{\varphi}-m^2\varphi\overline{\varphi}$, where the equations of motion are given by $(\Box+m^2)\varphi=0$;
 our results for global $U(1)$ symmetry can be straightforwardly generalized to the case of local $U(1)$ gauge symmetry. With the decompositions
for the quantum field and its conjugate momentum, 
\begin{eqnarray}
\hat{\varphi}(\vec{x},t)=\frac{1}{\sqrt{(2\pi)^D}}\int \frac{\diff\vec{k}}{w_{k}^{\frac{s}{2}}}\big[\hat{a}_{k} e^{-iw_{k}t}e^{i\vec{k}\cdot\vec{x}}+\hat{b}_{k}^{\dagger}   e^{iw_{k}t}e^{-i\vec{k}\cdot\vec{x}} \big];
\label{q}\\
\hat{\pi}(\vec{x},t)=\frac{i}{\sqrt{(2\pi)^D}}\int \frac{\diff\vec{k}}{w_{k}^{\frac{s}{2}-1}}\big[\hat{a}_{k}^{\dagger} e^{iw_{k}t}e^{-i\vec{k}\cdot\vec{x}}-\hat{b}_{k} e^{-iw_{k}t}e^{i\vec{k}\cdot\vec{x}} \big];
\label{p}
\end{eqnarray}
the equations of motion are satisfied as long as  the  (Lorentz invariant) dispersion relation $w_{k}^2=m^2+\vec{k}^2$ holds; the weight $s$ of the  measure $ \frac{\diff\vec{k}}{w_{k}^{\frac{s}{2}}} $  is in general a real quantity.
In order to gain insight into the physical meaning of a $s$-deformed measure, we can integrate the volume in the momenta space; we first notice that the integration of the usual Lorentz invariant measure $\int_{-\infty}^{\infty}\frac{\Diff{D}\vec{k}}{w_k}$ diverges for all dimensions, an anticipation of potential divergences.
At the level of the correlator the integration with a $s$-deformed measure will take the form $\int_{-\infty}^{\infty}\frac{\Diff{D}\vec{k}}{(m^2+k^2)^{s/2}}$, which we describe explicitly for different dimensions below; the table 1 shows that the volume in $k$-space is finite for all dimensions, and for all values of $s$, except for a divergence occurring at the limit $s\rightarrow D^+$, just at the dimension of the volume; from this divergence, the volume is decreasing monotonically to zero as $s\rightarrow+\infty$. This means that for each space-time $D+1$ dimension, the divergence occurs just at the dimension of the integration volume, in such a way that $s\rightarrow D+\epsilon$, with $\epsilon>0$.
Furthermore, in all cases the massless limit proves to be divergent at the region allowed, $s>D$; within this same region, all amplitudes approach to zero as $m\rightarrow+\infty$.

\begin{table}[h]
	\begin{center}
		\begin{tabular}{|c|c|c|c|}
			\hline
			&  $D=1$ & $D=2$ & $D=3$ \\ \hline
	\label{tab:p&q}
			$\int_{-\infty}^{\infty}\frac{\Diff{D}\vec{p}}{(m^2+k^2)^{s/2}}$ &  $\frac{\sqrt{\pi}\:\Gamma[\frac{s-1}{2}]}{m^{s-1}\Gamma[\frac{s}{2}]}$ & $\frac{\pi}{m^{s-2}(s-2)}$ & $\frac{\pi^{3/2}\:\Gamma[\frac{s-3}{2}]}{4m^{s-3}\Gamma[\frac{s}{2}]}$ \\ \hline
			Restrictions & $\Re[m^2]>0;\: s>1$ & $\Re[m^2]>0;\: s>2$ & $\text{Arg}[\frac{1}{m^2}]\leq\frac{2\pi}{3};\: \Re[m^2]>0;\: s>3$ \\ \hline
		\end{tabular}
	\end{center}
	\caption{Due to the behavior of the Gamma function with argument $\frac{s-D}{2}$, the volumes for $D=1,3$ diverge as $s\rightarrow D^+$, due to the restriction $s>D$; in general for $D$ odd, the volume is proportional to $\Gamma[\frac{s-D}{2}]$, which diverges as $s\rightarrow D^+$. For $D$ even, the volumes take the form $1/P(s)$, where $P(s)$ is a polynomial of the form $[s-D][s-(D-2)][s-(D-4)]\cdots[s-2]$, then they diverge also as $s\rightarrow D^+$; although the polynomial has as roots $s=D-2,D-4,\cdots,2$, the divergences are not realizable due to the restriction $s>D$. In general the argument for the complex mass takes the form $\text{Arg}[\frac{1}{m^2}]<\frac{2\pi}{D}$; in order to restrict to a real mass, we can choose $\text{Arg}[\frac{1}{m^2}]=0$ for all cases. Thus, the dependence on the mass takes the form $\frac{1}{m^{s-D}}$, and due to the restriction $s>D$ the volumes diverge in the mass-less limit $m\rightarrow 0$. Note that in all cases, selecting $s=D+1$, as the integer equal to the spacetime dimension guarantees convergence, although $s$ is not necessarily an integer.}
\end{table}

\newpage

The integrals (\ref{q}) and (\ref{p}) can also be regarded as a weighted Fourier transform \cite{fourier}, since the term $1/w_{k}^{s-1}\in L^{2}(\mathbb{R}^{D+1})$, as
\begin{equation}
\frac{1}{(2\pi)^{D+1}}\int\frac{d^{D+1}k}{(w_{k}^{s-1})^{2}}<\infty,
\end{equation} 
for all $s>(D+2)/2$. The appearance of the weight in the Fourier transform might suggest that the modification of the dispersion relation, originated from the Lorentz symmetry violation, exhibits anisotropy and non-uniform scaling within the modified spacetime structure. The origin of this generalization of the Fourier transform can, in principle, be determined by analyzing the Lorentz symmetry breaking using Hopf algebras and Quantum Groups \cite{hopf}. These approaches could provide us valuable insights into the modified or new symmetries of these systems.

\section{$s$-weighted propagators}
From the expressions (\ref{q}) and (\ref{p}) the subsequent amplitudes can be computed
\begin{eqnarray}
<0|\hat{\varphi}(\vec{x},t)\hat{\varphi}^{\dagger}(\vec{x'},t')|0>=\frac{\alpha}{2{(2\pi)^D}}\int \frac{d^D\vec{k}}{w_{k}^{s}} e^{-iw_{k}(t-t')}e^{i\vec{k}\cdot(\vec{x}-\vec{x'})}\equiv\Delta_{1}(x-x';s);
\label{prop1}\\
<0|\hat{\varphi}^{\dagger}(\vec{x'},t')\hat{\varphi}(\vec{x},t)|0>=\frac{\beta}{2{(2\pi)^D}}\int \frac{d^D\vec{k}}{w_{k}^{s}} e^{iw_{k}(t-t')}e^{-i\vec{k}\cdot(\vec{x}-\vec{x'})}\equiv \Delta_{2}(x-x';s);
\label{prop2}
\end{eqnarray}
where the following commutation relations for the annihilation and creation operators, along with the standard definition of the vacuum state, have been applied (for a detailed description on propagator field theory, see for example \cite{greiner}),
\begin{eqnarray}
[\hat{a}_{k},\hat{a}_{k'}^{\dagger} ]=\alpha\delta(k-k'),\quad [\hat{b}_{k},\hat{b}_{k'}^{\dagger} ]=\beta\delta(k-k');\qquad\qquad \hat{a}(k)\ket{0} =0= \hat{b}(k) \ket{0}\label{vacuum}.
\label{ac}
\end{eqnarray}
The constants $\alpha$ and $\beta$ introduce a mass scale (depending on the parameter $s$), as we will describe in the section \ref{oneplusone}. 

Furthermore, in the usual approach, one can make use of the Lorentz invariant measure
\begin{eqnarray}
\int d^4p \delta(p_{0}^2-\vec{k}^2-m^2)\Big |_{p_{0}>0}=\int \frac{\Diff{3} p}{2p_{0}}\Big |_{p_{0}=w_{k}},
\label{dirac1}
\end{eqnarray}
in order to express the propagator as a four-momentum integral;  in the case at hand, the s-deformed version of the above integral will lead to
the following $s$-deformed integrals;
\begin{eqnarray}
\Delta_{1}(x-x';s)=\frac{\alpha}{ {(2\pi)}^{D+1} }\int \frac{\Diff{{D+1}}{k}}{w_{k}^{s-1}} \frac{i}{k^2-m^2}e^{-ik_{0}(t-t')}e^{i\vec{k}\cdot(\vec{x}-\vec{x'})};
\label{delta1}\\
\Delta_{2}(x-x';s)=\frac{\beta}{ {(2\pi)}^{D+1} }\int \frac{\Diff{{D+1}}{k}}{w_{k}^{s-1}} \frac{i}{k^2-m^2}e^{ik_{0}(t-t')}e^{-i\vec{k}\cdot(\vec{x}-\vec{x'})};
\label{delta2}
\end{eqnarray}
thus, as we can observe, for $s\neq 1$ these expressions violate explicitly the Lorentz symmetry. The action of the operator $(\Box+m^2)$ on these amplitudes satisfy
\begin{eqnarray}
(\Box+m^2)\Delta_{1}(x-x';s)=-\frac{i\alpha}{ {(2\pi)}^{D+1} }\int \frac{\Diff{{D+1}}{k}}{w_{k}^{s-1}}e^{-ik_{0}(t-t')}e^{i\vec{k}\cdot(\vec{x}-\vec{x'})};
\label{Dalam1}
\end{eqnarray}
the source on the right hand side will reduce to the usual delta function for $s=1$; for $s>1$, the source proves to be a smoothed version of the singular delta function; specifically it can be rewritten in terms of a smooth space-like distribution, and a singular delta function in the time coordinate,
\begin{eqnarray}
\frac{1}{ {(2\pi)}^{D+1} }\int \frac{\Diff{{D+1}}{k}}{w_{k}^{s-1}}e^{-ik_{0}(t-t')}e^{i\vec{k}\cdot(\vec{x}-\vec{x'})}=\frac{1}{ {(2\pi)}^{D} }\int \frac{\Diff{D}{\vec{k}}}{w_{k}^{s-1}}e^{i\vec{k}\cdot(\vec{x}-\vec{x'})}\cdot\delta(t-t');
\end{eqnarray}
the space-like  integral has appeared previously in our approach \cite{ugalde} by calculating the field commutators (at the same time), showing convergence at short distances; in particular, we refer the reader to equation (14) and the accompanying table in \cite{ugalde} for the four-dimensional case. Such a table primarily represents the spatial component of the source in the above equation.

\section{$1+1$ quantum field theory}
\label{oneplusone}

By making use of the variable change $k=m\sinh(\theta)$, the expression (\ref{delta2}) can be rewritten as
\begin{eqnarray}\label{lc1}
	\Delta_{2}=\frac{\beta}{4\pi m^{s-1}} \int_{-\infty}^{\infty}\frac{e^{i\left[\frac{m(T+X)}{2}e^{-\theta} + \frac{m(T-X)}{2}e^{\theta}\right]}}{\cosh[s-1](\theta)}d\theta,
\end{eqnarray}
where $T=t-t'$, and $X=x-x'$; this expression captures the light-cone structure, and it will be convenient for our approximation.\footnote{In the approach at hand the parameters $\alpha$ and $\beta$ depend on $s$, in such a way that $[\alpha]=[\beta]=M^{s-1}$, where $M$ defines a mass scale; thus in the above expression the factor $\frac{\beta}{m^{s-1}}=\left(\frac{M}{m}\right)^{s-1}$ corresponds to a dimensionless parameter depending on $s$ \cite{ugalde}.}
For purely space-like separation, $T=0$, the commutator obtained from (\ref{lc1}) vanishes trivially for arbitrary values of $s$, $i\int_{-\infty}^{\infty}\sin(mX\sinh\theta)/\cosh^{s-1}\theta=0$, as wanted.
However, for purely time-like separation, $X=0$, the commutator does not vanish, since it takes the form $i\int_{-\infty}^{\infty}\sin(mX\cosh\theta)/\cosh^{s-1}\theta$ (as wanted).
In contrast, for purely space-like separation, the anti-commutator is defined by $\cos(X\sinh\theta)/\cosh^{s-1}\theta$, which does not vanish for arbitrary $s$, violating micro-causality (as established in the standard scheme).

Let us consider now the region of the light-cone defined by $T=X$, then, to construct an infini\-tesimal strip defined by the expansion $X+\delta X$, in such a way that for $\delta X>0$ we are located outside the light-cone, and for $\delta X<0$ inside it.

Hence, the first-order approximation in $\delta X$ for the expression (\ref{lc1}) will read
\begin{eqnarray}\label{lc2}
	\Delta_{2}(mX,m\delta X; s) =\frac{\beta}{4\pi m^{s-1}} \int_{-\infty}^{\infty}\frac{ e^{imXe^{-\theta}}}{\cosh^{s-1}\theta}[1-im\delta X\sinh\theta]\diff\theta,
\end{eqnarray}
the first term corresponds to the exact ($s$-deformed) amplitude on the light-cone, and the second one characterizes to the correction on the infinitesimal $\delta X$-strip near the light cone. Explicitly we have
\begin{eqnarray}
	\Delta_{2}(mX,m\delta X; s) =\frac{\beta}{4\pi m^{s-1}}\int_{-\infty}^{\infty}\diff\theta\left[\frac{\cos(mXe^{-\theta})+m\delta X\sinh(\theta)\cdot\sin(mXe^{-\theta})}{\cosh[s-1](\theta)}\right. \nonumber\\
	\left.+i\frac{\sin(mXe^{-\theta})-m\delta X\sinh(\theta)\cdot\cos(mXe^{-\theta})}{\cosh[s-1](\theta)}\right].
	\label{lc3}
	\end{eqnarray}

Since the two-points functions (\ref{prop1}) and (\ref{prop2}) are related by $\Delta_{1}=\overline{\Delta}_{2}$, with $\alpha=\beta$, then the expected value of the commutator $\bra{0} [\varphi,\varphi^+] \ket{0}$, and its correction are proportional to $\Delta_{2}-\overline{\Delta}_{2}$, the imaginary part of the above expression; similarly the anti-commutator $\bra{0} \{\varphi,\varphi^+\} \ket{0}$ and its correction are proportional to $\Delta_{2}+\overline{\Delta}_{2}$, the real part of the above expression. The expression (\ref{lc3}) can be determined explicitly in terms of the Gamma function and the hypergeometric functions $_1F_2$; for the anti-commutator we find out that

\begin{equation}\label{cone}
	\hspace*{-1.3cm}\Re[\Delta_{2}]=\frac{\beta}{m^{s-1}}\frac{1}{4\sqrt{\pi}}\left\{\frac{s\Gamma[\frac{s-1}{2}]}{2\Gamma[\frac{s}{2}]}\:_1F_2\left[\frac{s-1}{2};\frac{1}{2},\frac{3-s}{2};\frac{(mX)^2}{4}\right]+\frac{2^{s-2}\sqrt{\pi}\abs{mX}^{s-1}}{\Gamma[s]\cos(\frac{s\pi}{2})}\:_1F_2\left[s-1;\frac{s+1}{2},\frac{s}{2};\frac{(mX)^2}{4}\right]\right\}
\end{equation}
\begin{equation*}
	\hspace*{-0.3cm}\frac{\beta}{m^{s-1}}\frac{m\delta X}{4\sqrt{\pi}}mX\left\{\frac{\Gamma\left[\frac{s+1}{2}\right]}{(s-3)\Gamma\left[\frac{s}{2}\right]}\:_1F_2\left[\frac{s+1}{2};\frac{3}{2},\frac{5-s}{2};\frac{(mX)^2}{4}\right] -\frac{\Gamma\left[\frac{s-1}{2}\right]}{2\Gamma\left[\frac{s}{2}\right]}\:_1F_2\left[\frac{s-1}{2};\frac{3}{2},\frac{3-s}{2};\frac{(mX)^2}{4}\right]\right.
\end{equation*}
\begin{equation*}
	\hspace*{-0.5cm}\left.-\frac{2^{s-3}\sqrt{\pi}\abs{mX}^{s-3}}{\cos(\frac{s\pi}{2})\Gamma[s-1]}\left(\:_1F_2\left[s-1;\frac{s-1}{2},\frac{s}{2};\frac{(mX)^2}{4}\right]+\frac{(mX)^2}{s(s-1)}\:_1F_2\left[s-1;\frac{s+1}{2},\frac{s}{2}+1;\frac{(mX)^2}{4}\right]\right)\right\},
\end{equation*}
and for the commutator
\begin{equation}\label{ccone}
	\hspace*{-1.3cm}\Im[\Delta_{2}]=\frac{\beta}{m^{s-1}}\frac{mX}{4\sqrt{\pi}}\left\{\frac{\Gamma[\frac{s-2}{2}]}{\Gamma[\frac{s-1}{2}]}\:_1F_2\left[\frac{s}{2};\frac{3}{2},2-\frac{s}{2};\frac{(mX)^2}{4}\right]+\frac{2^{s-2}\sqrt{\pi}\abs{mX}^{s-2}}{\Gamma[s]\sin(\frac{s\pi}{2})}\:_1F_2\left[s-1;\frac{s+1}{2},\frac{s}{2};\frac{(mX)^2}{4}\right]\right\}
\end{equation}
\begin{equation*}
	\hspace*{-0.3cm}+\frac{\beta}{m^{s-1}}\frac{m\delta X}{4\sqrt{\pi}}\left\{\frac{\Gamma\left[\frac{s-2}{2}\right]}{2\Gamma\left[\frac{s-1}{2}\right]}\:_1F_2\left[\frac{s}{2};\frac{3}{2},2-\frac{s}{2};\frac{(mX)^2}{4}\right] -\frac{\Gamma\left[\frac{s-1}{2}\right]}{2\Gamma\left[\frac{s}{2}\right]}\:_1F_2\left[\frac{s-1}{2};\frac{1}{2},1-\frac{s}{2};\frac{(mX)^2}{4}\right]\right.
\end{equation*}
\begin{equation}\label{cccone}
	\hspace*{-0.5cm}\left.-\frac{2^{s-3}\sqrt{\pi}\abs{mX}^{s-2}}{\sin(\frac{s\pi}{2})\Gamma[s-1]}\left(\:_1F_2\left[s-1;\frac{s-1}{2},\frac{s}{2};\frac{(mX)^2}{4}\right]+\frac{(mX)^2}{s(s-1)}\:_1F_2\left[s-1;\frac{s+1}{2},\frac{s}{2}+1;\frac{(mX)^2}{4}\right]\right)\right\}.
\end{equation}

We will focus on the commutators and their corrections, as they respect microcausality; moreover, we can observe that the exact expression on the light-cone (\ref{ccone}) is odd under $X\rightarrow-X$, whereas the correction (\ref{cccone}) is even under the same transformation; however, by imposing time ordering on the propagator on the light-cone, this discrete symmetry will be broken, eliminating the regions with dominating negative amplitudes. 

\begin{figure}[H]
	\begin{center}
		\includegraphics[width=.9\textwidth]{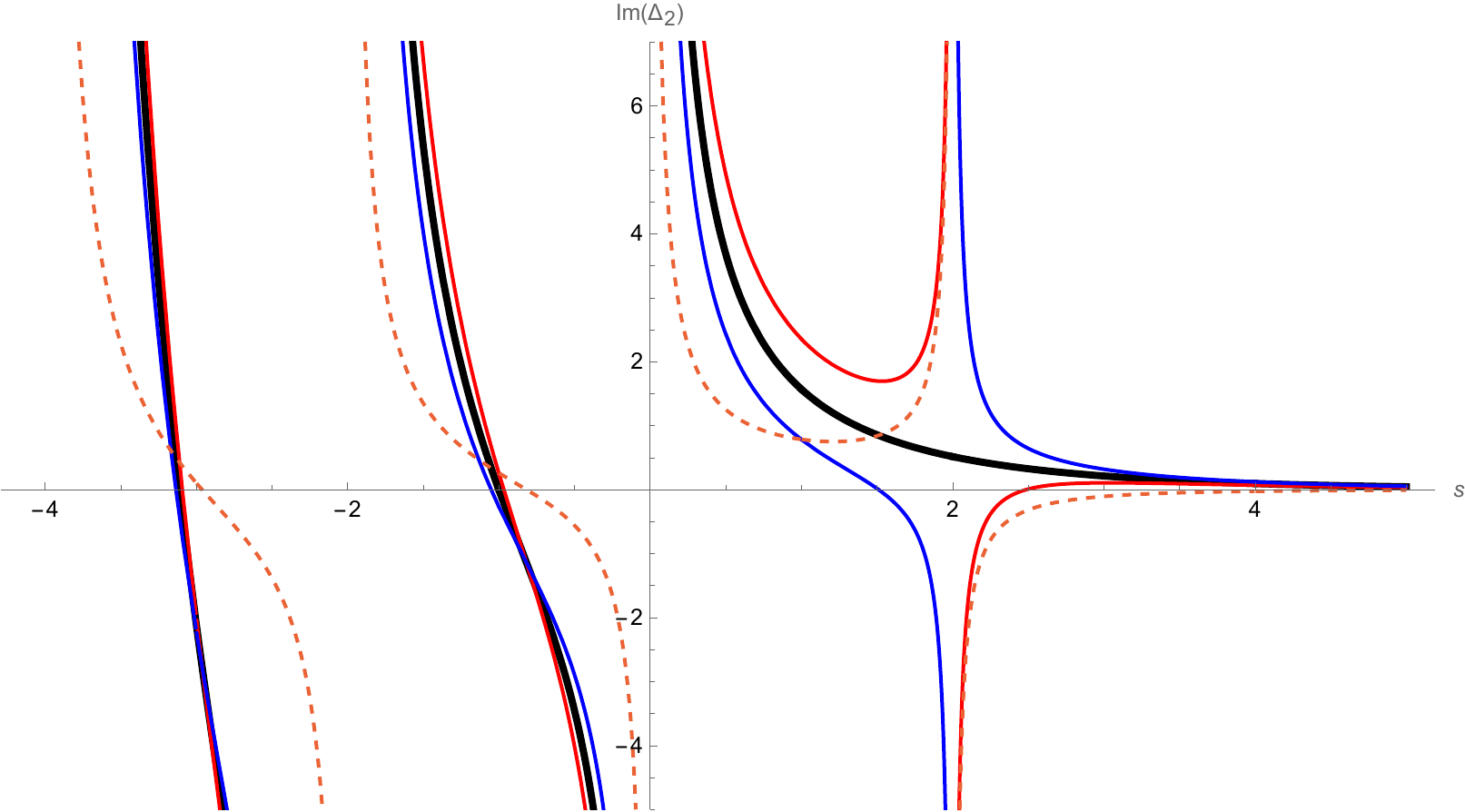}
		\caption{The mass scales and space-time separations are fixed at $mX=2$, $m\abs{\delta X}=0.5$, and $\frac{M}{m}=0.5$; in the range $s<2$, the amplitude outside (red) the light-cone is greater than the amplitude on (black), and inside (blue) the light-cone, which does not respect the classical hierarchy . Hence, the range of interest will be $s>2$, where the relationship outside, on, and inside the light-cone is as expected classically; the inside-amplitude is greater than the on-amplitude, and the later is greater than the outside-amplitude. Note that the exact amplitude on the light-cone is well defined for $s=2$. All amplitudes go to zero as $s\rightarrow+\infty$, where the condition $\frac{M}{m}<1$ plays a relevant role. This case corresponds qualitatively to the strong interaction scenario.}
		\label{prop less1}
	\end{center}
\end{figure}

\begin{figure}[H]
	\begin{center}
		\includegraphics[width=.9\textwidth]{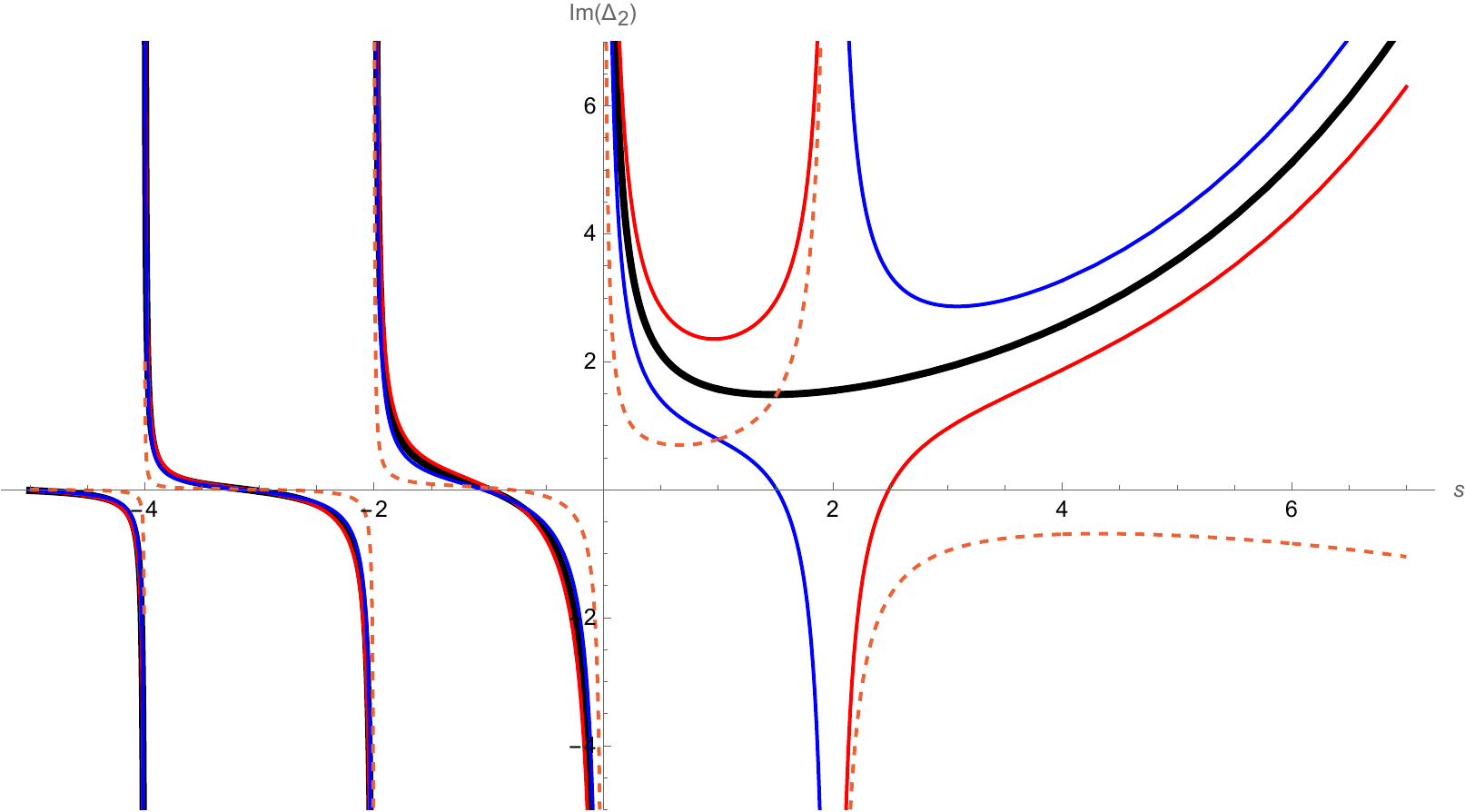}
		\caption{We consider now $\frac{M}{m}>1$, with $\frac{M}{m}=1.5$, and with the same values for $mX$ and $m\abs{\delta X}$ considered in the figure \ref{prop less1}; the condition $\frac{M}{m}>1$ will induce the divergence of all the amplitudes as $s\rightarrow+\infty$, as opposed to the case described in the figure \ref{prop less1}, with a vanishing limit for all the amplitudes. Again, the region of interest is for $s>2$; note that the inside-amplitude (blue curve) has an absolute minimum for certain $s\approx3$. This case corresponds qualitatively to the weak interaction scenario.} 
		\label{prop greater1}
	\end{center}
\end{figure}

Now, we describe first the dependence on the deformation parameter $s$, by fixing the mass scales and space-time separations. For large $s$, the amplitude will depend sensitively on the relation between the mass scales; in the figures \ref{prop less1} and \ref{prop greater1}, the exact amplitude on the light-cone (\ref{ccone}) corresponds to the black curve, with divergences at $s=0,-2,-4,\cdots$; the correction (\ref{cccone}) corresponds to the dashed curve with divergences at $s=2,0,-2,-4,\cdots$. The full expression, namely, exact amplitude plus its correction corresponds to the red curve for the region outside the light-cone ($\delta X>0$), and to the blue curve for the region inside the light-cone ($\delta X<0$). In the figures it is noted that although the exact amplitude on the light cone is finite for $s=2$ (just the background dimension), the $\delta X$-correction diverges; this can be considered as a sign of instability of the exact amplitude under small perturbations on the light-cone. This instability can be associated with the UV momenta contributions, since the infrared regime shows stability precisely for $s=2$ (see section \ref{IRregime}).  As we shall see, the stability can be established in the full description under the restriction $s>2$.

Our description is sensitive on the ratio of the mass scales through the global factor $\beta/m^{s-1}=(M/m)^{s-1}$, since a bifurcation emerges from the inequalities $M/m<1$, or $M/m>1$. The particle physics provides specific scenarios as follows: considering the mass scale defined by the strong interaction with $M=1 Gev$, and a mass for a Higgs-like field for our model,
identified  with the mass determined experimentally for the Higgs particle $m=125 Gev$, the first scenario is achieved with $M/m=8\times 10^{-3}$. Moreover, the second scenario can be achieved by considering the electroweak scale with $M=246 Gev$, and the same Higgs mass, hence $M/m\approx 2$. Another case for the second scenario ($M/m>1$) is at the Planck scale with $M\approx 10^{19}Gev$ (for more details on the particle physics numerology in the approach at hand, see section 6 in \cite{ugalde}). 
Due to the qualitative behavior of the amplitudes depends on these inequalities, one can retain in mind these specific particle physics scenarios along the paper; 
the two scenarios for the commutators are described in the figures below.

\begin{figure}[H]
	\begin{center}
		\includegraphics[width=.82\textwidth]{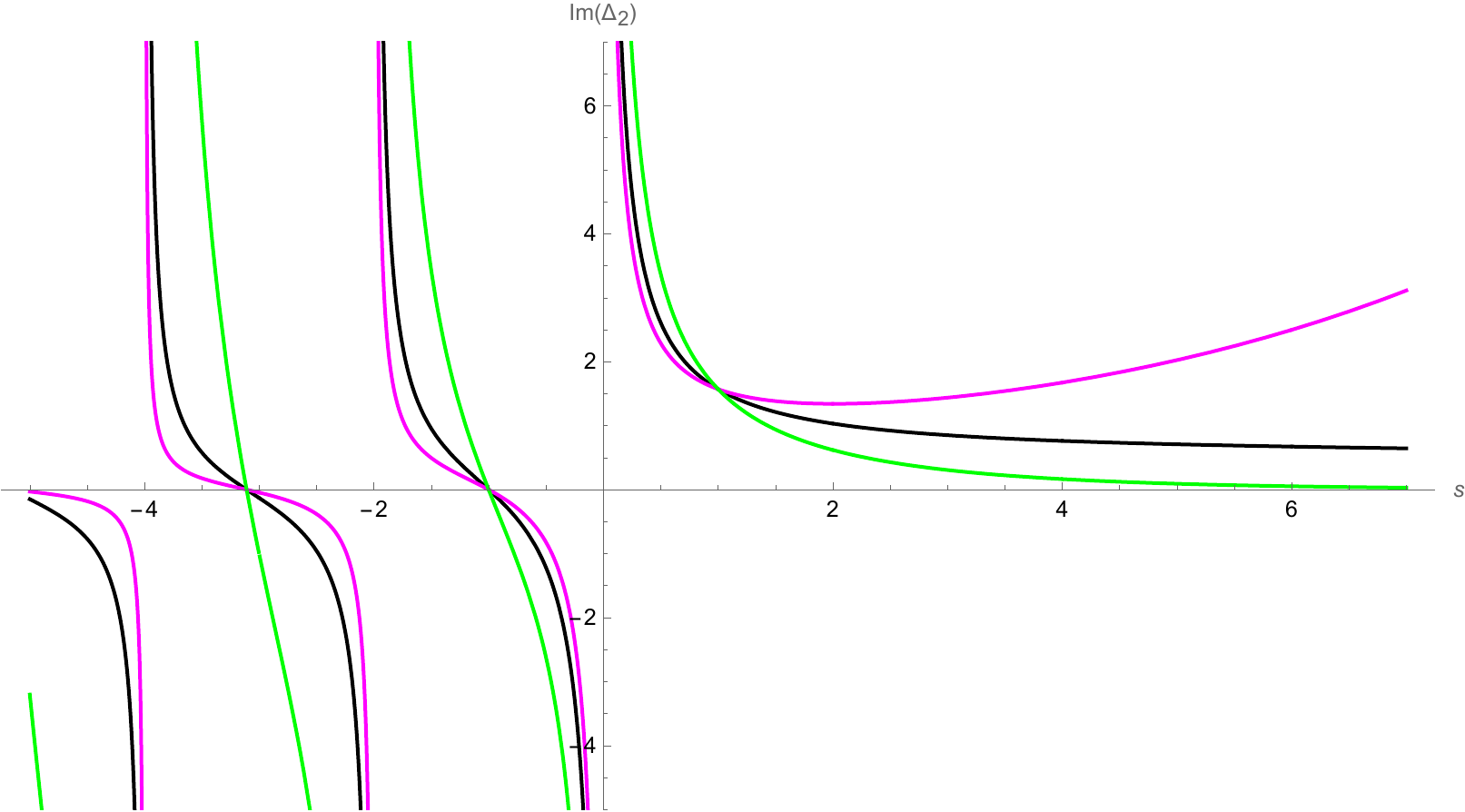}
		\caption{We describe here the exact amplitudes on the light-cone as function on $s$ with $mX=2$. The case $\frac{M}{m}=1$ is represented by the black curve; the case $\frac{M}{m}=1.3$ corresponds to the magenta curve (weak interaction scenario), and $\frac{M}{m}=0.6$ to the green curve (strong interaction scenario). The amplitudes are finite as $s\rightarrow+\infty$ for $\frac{M}{m}\leq1$; for $\frac{M}{m}=1$ such a limit is nonvanishing, as opposed to the case described in the figure \ref{prop less1}. Note that for these amplitudes the region of interest is $s>0$; in this region the on-amplitudes are positive definite.} 
		\label{Mm light11}
	\end{center}
\end{figure}

\vspace*{-0.3cm}
We describe now the amplitudes as functions on $mX$, for fixed values of $s$; as we shall see, for  integer values of $s$ the amplitudes can be described alternatively in terms of the exponential and exponential integral functions. In the table below, the column $A)$ describes the anti-commutator for different values of $s$; its $\delta X$-correction is described correspondingly by the column $B)$. Similarly the columns $C)$ and $D)$ describe the field commutator and its $\delta X$-correction respectively. The divergence for the $\delta X$-correction 	in the case $s=2$ in the column D) for the commutator, can be associated with the UV contributions for the momenta, since in the infrared regime described in the section 5, such a divergence will disappear.

\renewcommand{\arraystretch}{1.3}
\begin{table}[h]\label{tab:lc3}
		\begin{tabular}{|c|c|c|}
			\hline
			&  Anti-commutator: A) $\int_{-\infty}^{\infty}\frac{\cos(mXe^{-\theta})}{\cosh[s-1](\theta)}\diff\theta$ & B) $\:-\int_{-\infty}^{\infty}\frac{\sinh(\theta)\cdot\sin(mXe^{-\theta})}{\cosh[s-1](\theta)}\diff\theta$ \\ \hline
			$s=2$ &  $\pi e^{-\abs{mX}}$ & $\pi\text{ sgn}(mX)\left[e^{-\abs{mX}}-\frac{1}{2}\right]$ \\ \hline
			$s=3$ & $\hspace*{0.032cm}mX\left[e^{mX}Ei(-mX)-e^{-mX}Ei(mX)\right]+2\hspace*{0.032cm}$ & $\hspace*{0.032cm}-mX\left[e^{mX}Ei(-mX)+e^{-mX}Ei(mX)\right]\hspace*{0.032cm}$ \\ \hline
			$s=4$ & $\frac{\pi}{2}\left[1+\abs{mX}-(mX)^2\right]e^{-\abs{mX}}$ & $-\frac{\pi}{2}mX(\abs{mX}-1)e^{-\abs{mX}}$ \\ \hline
		\end{tabular}
\end{table}
\vspace*{-0.5cm}
\begin{table}[h]
		\begin{tabular}{|c|c|c|}
			\hline
			& Commutator: C) $\int_{-\infty}^{\infty}\frac{\sin(mXe^{-\theta})}{\cosh[s-1](\theta)}\diff\theta$ & D) $\:-\int_{-\infty}^{\infty}\frac{\sinh(\theta)\cdot\cos(mXe^{=\theta})}{\cosh[s-1](\theta)}\diff\theta$\\ \hline
			$s=2$ & $e^{-mX}Ei(mX)-e^{mX}Ei(-mX)$ & $\begin{array}{l} \infty\text{, if }\theta\in(-\infty,0); \\ \text{finite, if }\theta\in(0,+\infty)\end{array}$ \\ \hline
			$s=3$ & $\pi mXe^{-\abs{mX}}$ & $-\pi\abs{mX}e^{-\abs{mX}}$ \\ \hline
			$s=4$ & $\begin{array}{c} \frac{1}{2}\left[1+mX-(mX)^2\right]e^{-mX}Ei(mX) \\ +\frac{1}{2}\left[-1+mX+(mX)^2\right]e^{mX}Ei(-mX)+mX\end{array}$ & $\begin{array}{c} \frac{1}{2}mX\left[(mX-1)e^{-mX}Ei(mX)\right. \\ \left.+(mX+1)e^{mX}Ei(-mX)\right]\end{array}$ \\ \hline
			\end{tabular}
			\caption{$Ei$ is the exponential integral; in the column for $s=1$ (the unbroken Lorentz symmetry case), the amplitude $C)$ reduces to a constant, namely $\int_{-\infty}^{\infty}\sin(mXe^\theta)=\frac{\abs{mX}}{mX}\frac{\pi}{2}$ (which is not physically admissible), and its correction $D)$ diverges. For $s=5$ the amplitude $C)$ reads $\frac{\pi}{6}mXe^{-\abs{mX}}\left[3+3\abs{mX}-(mX)^2\right]$, and for $s=7$ it reads $\frac{\pi}{120}mXe^{-\abs{mX}}\left\{45+45\abs{mX}+(mX)^2\left[(mX)^2-10\abs{mX}+5\right]\right\}$. The commutator $C)$ is odd and the correction $D)$ is even under $X\rightarrow-X$; these symmetries are interchanged for the anti-commutator $A)$ and the correction $B)$.}
\end{table}

\newpage
\section{The Feynman propagators on and near the light cone}

From this point the amplitudes are described by completeness without time-ordering; a time-ordered propagator can be obtained from $\Delta_{2}$ by considering the restriction $T>0$; it means that the s-weighted Feynman propagators are described in the figures by the part with $X\geq 0$, with the original odd discrete symmetry ($X\rightarrow -X$) broken.
In this region the Feynman propagators in the vicinity of the light cone will be positive definite for certain range of the weight $s$; although in other range for the weight $s$ such a propagator will retain regions with negative values, we discuss on the possibility 
that such amplitudes may be admisible in the description (see section \ref{negative}).

The amplitudes and their corrections are illustrated in the figures below as functions on the dimensionless variable $mX$; 
for a fixed non-vanishing mass, these figures can be interpreted also as the profiles for the amplitudes in function of the separation $x'-x$. As a result of the presence of the global mass factor $1/m^{s-1}$ in the expression (\ref{lc3}), the amplitudes as functions on the mass show qualitatively different profiles, without critical points, such as maxima and minima; in fact all amplitudes diverge for a massless field; additionally the amplitudes go to zero as $m\rightarrow\infty$. Note however that the amplitudes are always finite for a massive field on the light cone for $s>1$; meanwhile, for the unbroken Lorentz symmetry case, $s=1$, the amplitude for  a massive field diverges on the light cone.

\begin{figure}[H]
  \begin{center}
    \includegraphics[width=.35\textwidth]{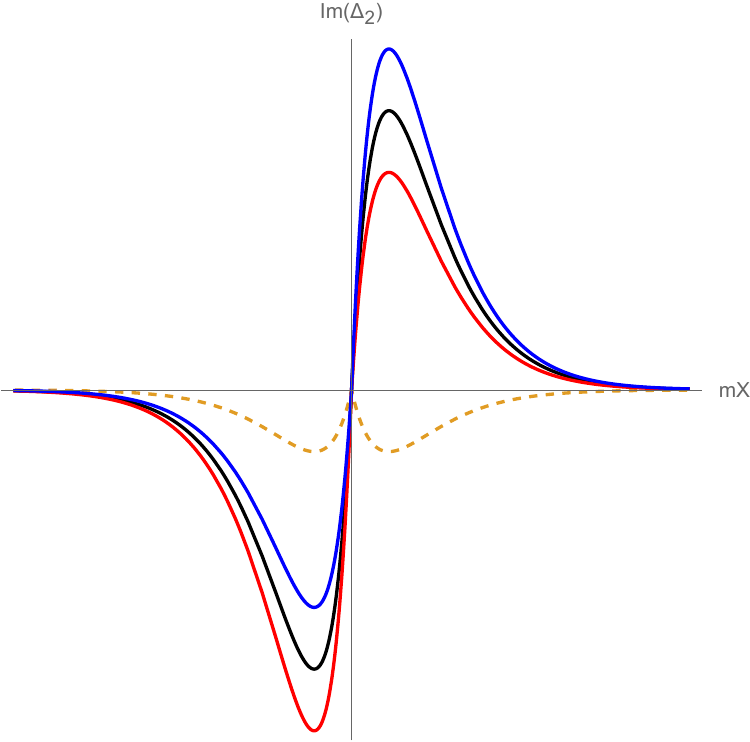}
  \caption{The propagator C) on the light cone for $s=3$ corresponds to the black curve; its correction D) is represented by the dashed curve, with negative values in the full range; the amplitude plus its correction is represented by the blue curve for $\delta X<0$ (inside the light cone), and by the red curve for $\delta X>0$ (outside the light cone); we have considered that $m>0$, and that $|\delta X|=0.2$. Note that the amplitude inside the light cone is always bigger than that on the light cone; similarly the amplitude outside the light cone is smaller than that on the light cone: the classical hierarchy is respected. All amplitudes are finite in the full range; additionally all amplitudes go to zero as $X$ goes to $\infty$.} 
   \label{propa1}
  \end{center}
\end{figure}

The amplitude C) and its correction D) for $s=3$ are illustrated in the figure (\ref{propa1}); the correction D)  has the minima at $|X|=1$, it means for $|x'-x|=\frac{1}{m}$, just at the Compton longitude; coincidently the critical points (in particular the maxima) for the amplitudes are located just at the same point; note that while $|x'-x|<\frac{1}{m}$ the amplitudes are increasing; after crossing the maximum at the Compton longitude, all amplitudes are decreasing. The curves with their critical points can be explained as a sort a competition between the spatial separation $x'-x$ and the Compton longitude. In this case the Feynman propagator ($X>0$) on the light cone and their $\delta X$-corrections are positive definite in the full range.

\begin{figure}[H]
  \begin{center}
    \includegraphics[width=.35\textwidth]{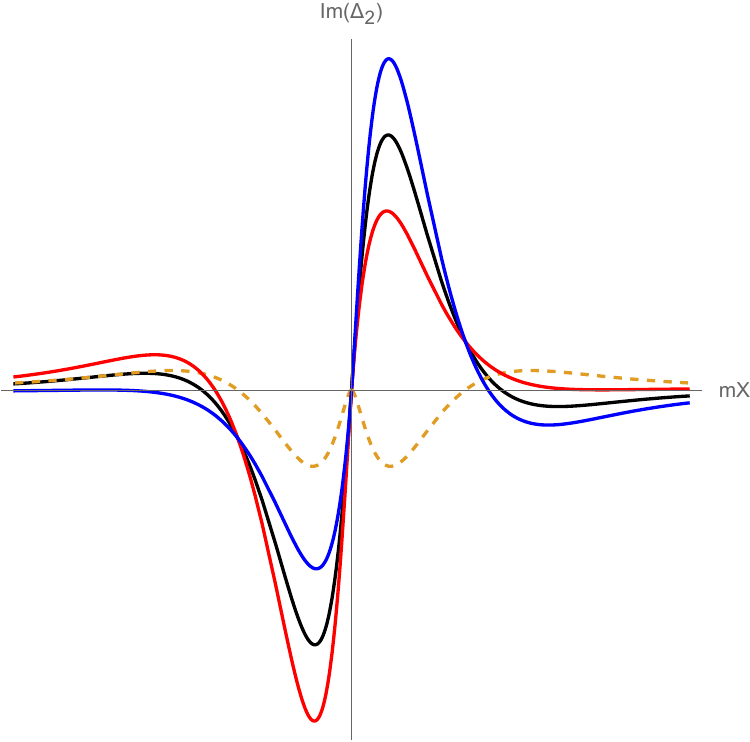}
  \caption{The amplitude C) with its corrections for $s=4$ with the same color code for the light cone regions; as opposed to the previous case with $s=3$, the curves have crossing points; a relevant cross point corresponds to the zero for the correction located at $mX\approx3.38$, with a separation around three times the Compton longitude; at this point the amplitudes on, inside, and outside the light cone coincide to each other; after crossing this point, the roles of greater/smaller amplitudes are interchanged. Again, all amplitudes are finite in the full range, and all of them go to zero as $X$ goes to $\infty$.} 
   \label{propa2}
  \end{center}
\end{figure}

The amplitude $C)$ and its correction for $s=4$, is described in the figure $\ref{propa2}$; the correction $D)$ (a transcendental expression) has the minimum at $m|X|\approx 1.153$, thus,  $|x'-x| \approx \frac{1}{m}+0.153 \frac{1}{m}$, it corresponds to the Compton longitude, with a  correction on the order $10^{-1}$. Note that as opposed to the previous case with $s=3$, in this case the amplitudes on and near the light cone take negative values in certain region; in fact the roots will appear in the strict range $s>3$, as we shall describe below by using the ceiling function.

In the above figures we realize that a small deviation $\delta X$ from the light cone, implies a small perturbation of the exact amplitudes on the light cone, and depending on the sign of that deviation, there exists a correspondence of the perturbed amplitude with the classical propagation inside, on, and outside the light cone. Moreover, this correspondence {\it small deviation/small deformation} can be interpreted as a proof for the stability of the $s$-weighted quantum propagation of a massive particle on the light cone. As we shall see, a similar interpretation can be given for the stability of the propagators on the $s$-space. 

\section{Propagators in the $s$-space and the flattening effect}
\begin{figure}[h!]
	\centering
	\begin{minipage}{0.31\textwidth}
		\centering
		\includegraphics[width=0.95\textwidth]{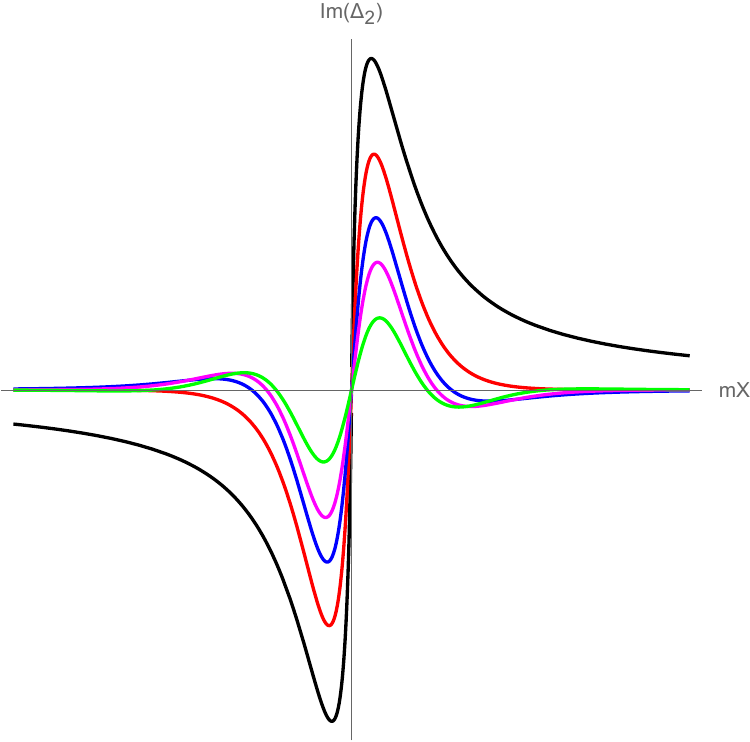}
	\end{minipage}%
	\hspace{5mm}
	\begin{minipage}{0.31\textwidth}
		\centering
		\includegraphics[width=0.95\textwidth]{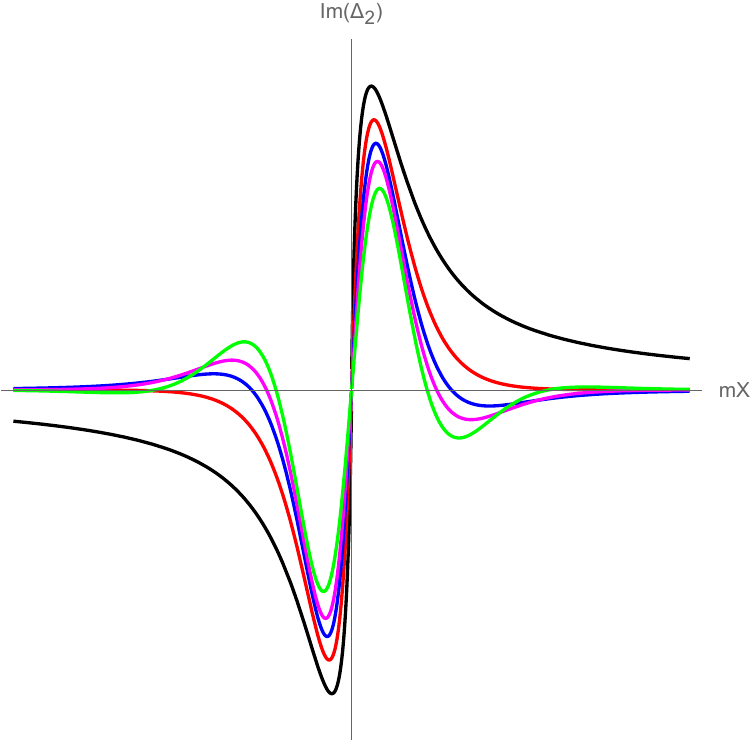}
	\end{minipage}
	\begin{minipage}{0.31\textwidth}
		\centering
		\includegraphics[width=0.95\textwidth]{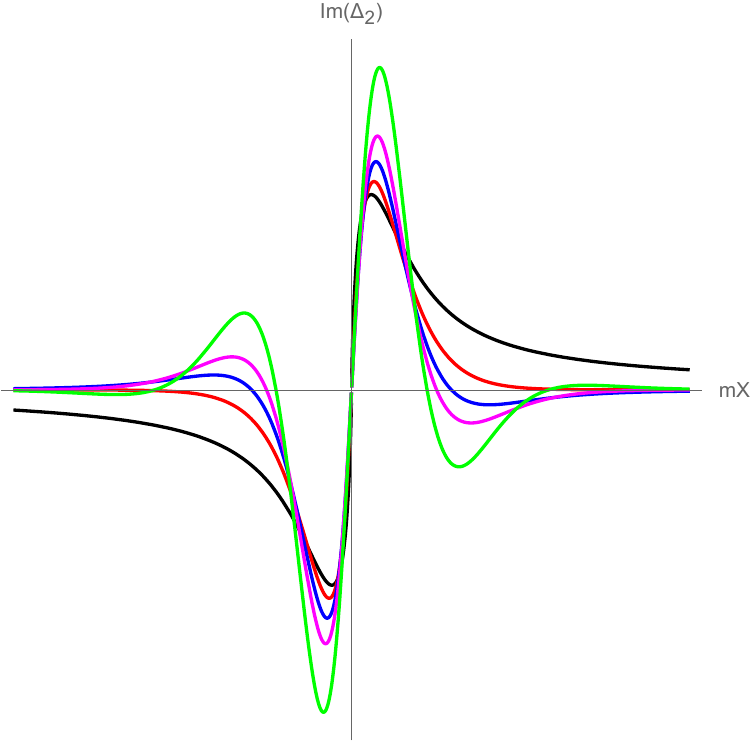}
	\end{minipage}
	\caption{The exact amplitudes C) on the light cone, for $s=2,3,4,5,7$ with the color code, $2$-black, 3-red, 4-blue, 5-magenta, and 7-green; to the left we have $\frac{M}{m}<1$, with $\frac{M}{m}=0.8$, at the center $\frac{M}{m}=1$, and to the right $\frac{M}{m}>1$, with $\frac{M}{m}=1.2$. To the left the curves are flattening as $s\rightarrow+\infty$ (qualitatively the strong interaction scenario); at the center for $mX>0$, although the maxima go to zero as $s\rightarrow+\infty$, the minima are increasing their (absolute) values as $s$ is increasing. To the right the curves do not flatten (the weak interaction scenario, or Planck scale scenario), since maxima and minima are increasing their values as $s\rightarrow+\infty$. Note that all the amplitudes are finite in the full range and go to zero as $mX$ goes to $\infty$.}
	\label{propa3} 
\end{figure}

In the figures \ref{propa1} and \ref{propa2} the commutators near the light-cone have been described by certain fixed values of $s$; we compare now the commutators $C)$ for different values of $s$; due to the expressions for the odd values $s=3,5,7$ are relatively easier to hand (since they only require the exponential function and polynomials), we illustrate in the figure (\ref{propa3}), the corresponding functions on $mX$ in order to gain insight in the behavior for increasing $s$. For odd values with $s\geq 9$, the functions require additionally of the exponential integral, in similarity with the expressions $C)$ and $D)$ with $s=4$ in the table above.

Note that in the three cases illustrated in the figure \ref{propa3}, the absolute maxima are located always around the Compton wavelength, with  small corrections as $s$ is increasing; for example, for the case $s=3$ the (absolute) maximum is located exactly at $m|X|=1$, and for the case $s=4$  such a location admits a $10^{-1}$-order correction, as stated previously; this fact is also valid for the corresponding $\delta X$-corrections, which have their maxima with small deviations respect to the maxima of the exact amplitudes, as it is illustrated in the figures \ref{propa1} and \ref{propa2}. This observation allows us to establish that the decreasing (increasing) of the amplitudes and their corrections in regions smaller (greater) than a distance around the Compton wavelength, is an universal property for the  $s$-weighted (Feynman) propagators.

The behavior of the propagators in the $s$-space described in the figures \ref{propa3} can be understood from a glance at the full expression for the $s$-depending measure in the equation (\ref{lc1}), which must satisfy
\begin{equation}\label{sinfty}
	\lim_{s\rightarrow +\infty}\left(\frac{1}{\frac{m}{M}\cosh\theta}\right)^{s-1}=0,
\end{equation}
in order to obtain the flattening for the propagators; a fully flattened propagator will imply that a quantum propagation does not exist in such a limit, which will be achieved only in the strong interaction scenario.
The above limit requires that
\begin{equation}\label{sinfty1}
	\ln(\frac{m}{M}\cosh\theta)>0;\qquad\frac{k}{m}=\sinh\theta;
\end{equation}
however, this condition is not satisfied automatically; for $\frac{M}{m}=1$, $\ln(\cosh\theta)=0$ for $k=0$, thus the flattening is not achieved due to effects of the zero-point for the momenta  (central-figure). For $\frac{M}{m}<1$, the condition (\ref{sinfty1}) is satisfied for all $k$; thus the flattening is achieved (left-figure). The case $\frac{M}{m}>1$ is subtler; the condition (\ref{sinfty1}) is not satisfied in a finite interval around the zero-point 
\begin{equation}\label{sinfty2}
	\abs{k}\leq m\sqrt{\frac{M}{m}-1}\sqrt{\frac{M}{m}+1} ; \qquad\qquad \frac{M}{m}\geq1
\end{equation}
thus, the contribution of momenta in this interval spoils the flattening in the right-figure; note that the case $\frac{M}{m}=1$ is included in this inequality, reducing the finite interval to the zero-point. Therefore, in relation to the particle physics scenarios, the non-flatenning effect on the the amplitudes in the $s$-space is relevant only at the weak interaction or Planck scale scenarios,  and it is not present in the strong interaction scenario.

\begin{figure}[h!]
	\centering
	\begin{minipage}{0.31\textwidth}
		\centering
		\includegraphics[width=0.95\textwidth]{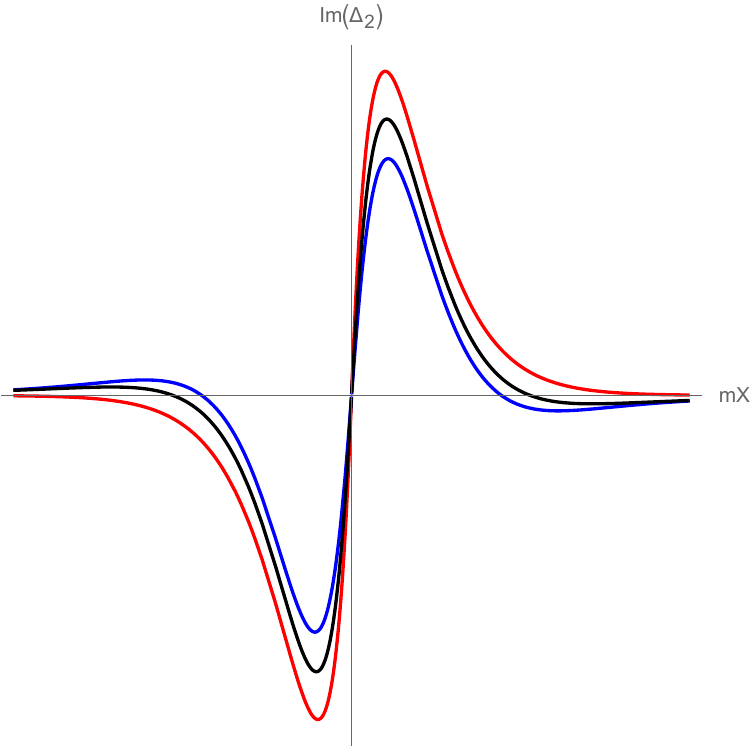}
	\end{minipage}%
	\hspace{5mm}
	\begin{minipage}{0.31\textwidth}
		\centering
		\includegraphics[width=0.95\textwidth]{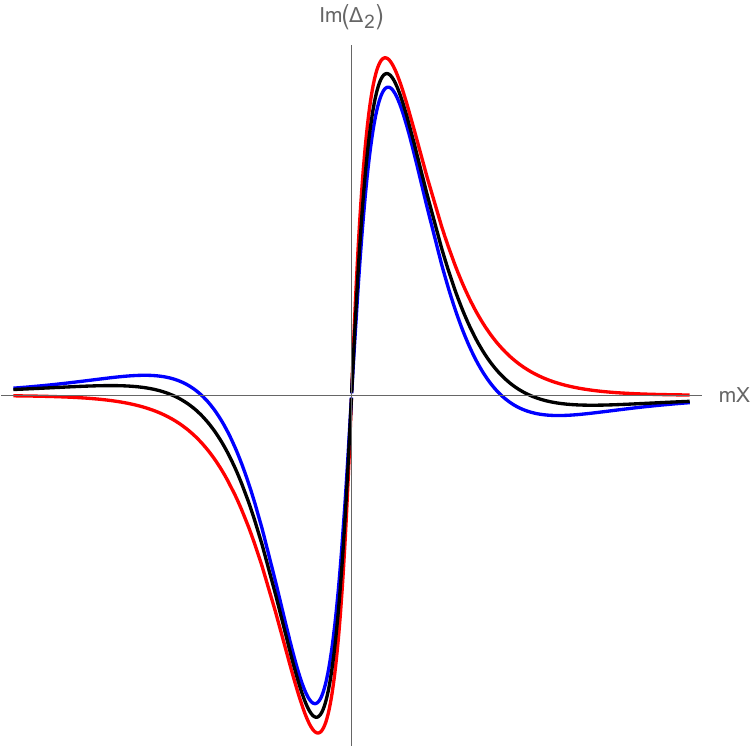}
	\end{minipage}
	\begin{minipage}{0.31\textwidth}
		\centering
		\includegraphics[width=0.95\textwidth]{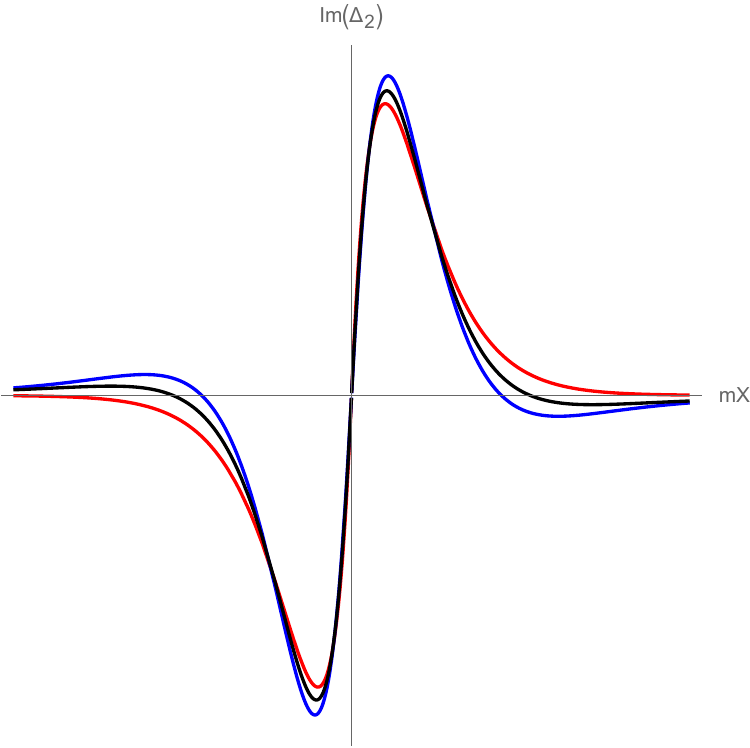}
	\end{minipage}
	\caption{The exact amplitudes (\ref{ccone}) on the light cone, for 3-red, and 4-blue, and the black curve represents the amplitude on the light cone for an intermediate fractional value, $s=3.5$. This shows that the amplitude for $s=3.5$ will lie between $s=3$ and $s=4$, and therefore it will have the same number of local maxima, local minima, and the same number of roots as $s=4$. To the left we have $\frac{M}{m}<1$, with $\frac{M}{m}=0.8$, at the center $\frac{M}{m}=1$, and to the right $\frac{M}{m}>1$, with $\frac{M}{m}=1.2$. Note that all the amplitudes are finite in the full range and go to zero as $mX$ goes to $\infty$.}
	\label{propa4} 
\end{figure}

Up to this point we have analyzed only integer values of $s$; however the amplitude (\ref{ccone}) for any $s$ between its two nearest integers, will lie between the amplitudes for those integers. In fact, the number of roots of the exact amplitude on the light-cone for any $s>2$, will be exactly $\ceil{s}-3$, where $\ceil{s}$ is the ceiling function; this intercalation of amplitudes is illustrated in the figure \ref{propa4}.

The intercalation of the amplitudes as $s$ is increasing (with possible intersections and regions with negative values) can be understood as a stability criterion for the propagators in $s$-space, since a small change in the real parameter $s$ will imply a small change in the amplitudes, regardless the value of the ratio $M/m$. As we shall see, this stability on the $s$-space will persist in the infrared regimen described below.

\section{On negative values for the amplitudes}
\label{negative}
As we have seen, the Feynman propagators with time ordering retain regions with negative amplitudes for $s>3$; if we restrict ourselves to positive definite amplitudes, then we must to impose the restriction $2<s\leq3$; within this range the Feynman propagators on the light-cone, and their $\delta X$-corrections are positive definite with a physically well defined asymptotic limit for large argument. This criterion may be used, in fact, as the first restriction on the values of the real parameter $s$, in order to obtain a specific finite QFT with a particular value for $s$. The restriction has a physical meaning with respect to the two dimensional background, since it corresponds to $D+1<s\leq D+2$.

However, if one wants to incorporate the amplitudes with regions with negative values for $s>3$, then that is plausible for $\frac{M}{m}<1$, since such values are only small fluctuations that go to zero as the amplitudes undergo the flattening with $s$ increasing (figure \ref{propa3} to the left); thus, one can to choose a convenient $s$ for suppressing the fluctuations, which can be as small as wanted. Note that this is not possible for the cases with $\frac{M}{m}\geq1$ (figure \ref{propa3} to the center and to the right), since maxima and/or minima are increasing their (absolute) values.

On the other hand, one can eliminate the regions with negative values for $s>3$ by considering that the present approach is not valid beyond the first root for the amplitudes of their $\delta X$-corrections; for example for the case $s=4$ (see figure \ref{propa2}), the (only) roots for the inside, on, and outside amplitudes are in order, $mX\approx4.07$, $mX\approx4.45$ and $mX\approx6.481$; hence all the amplitudes are positive definite for $mX\leq4.07$, which means for a separation $|x'-x|$ smaller than four times the Compton wavelength.

\section{Infrared limit for the case $1+1$ QFT}
\label{IRregime}

The non-flattening interval (\ref{sinfty2}) around the zero-point, can be considered as the infrared regimen defined by the mass scales; in particular one can consider $\frac{M}{m}=\sqrt{2}$ as a simple realization of the inequality (qualitatively the weak interaction scenario, or the Planck scale scenario), and to restrict the analysis by letting $k$ be smaller than $m$; hence, the non-flattening will be manifest as $s$ is increasing. In this regimen we shall be able to describe the propagation on the entire $1+1$ background; the propagator is physically well defined on the different regions only for certain values of $s$. In particular a well defined propagator on the light-cone will allow us to identify the weight $s$ exactly with the background dimension;
this identification was anticipated in the discussion on the convergence of the $s$-weighted volume in the momenta space (see the table 1).
By considering the expansions,
\begin{equation}\label{ApprSqrt IR 1}
	\frac{1}{\sqrt{m^2+k^2}^s}\approx m^{-s}\left(1-s\frac{k^2}{2m^2}\right),
\end{equation}
\begin{equation}\label{ApprExp IR 1}
	e^{i\sqrt{m^2+k^2}T}\approx[\cos(mT)-\frac{k^2}{2m}T\sin(mT)]+i[\sin(mT)+\frac{k^2}{2m}T\cos(mT)],
\end{equation}
the expression (\ref{prop2}) becomes
\begin{equation}\label{IR1}
	\hspace{-0.5cm}\Delta_{2}= \frac{\beta}{4\pi m^{s-1}}\int_{-1}^{1}(1-\frac{s}{2}\omega^2)\left([\cos(mT)-\frac{\omega^2}{2}mT\sin(mT)]+i[\sin(mT)+\frac{\omega^2}{2}mT\cos(mT)]\right)\cos{(mX\omega)}\diff\omega,
\end{equation}	
where $\omega\equiv k/m$. Explicitly the real part for the anti-commutator will read \\
\begin{equation}\label{ReR}
	\Re[\Delta_{2}]=\frac{\beta}{8\pi m^{s-1} (mX)^{4}}\biggr[\frac{T}{X}\sin(mT)\Bigr\{4mX[(mX)^2(s-1)-6s]\cos(mX)
\end{equation}
\vspace*{-0.6cm}
\begin{equation*}
	\hspace{-0.9cm}+[(mX)^4(s-2)-4(mX)^2(3s-1)+24s]\sin(mX)\Bigr\}-2mX\cos(mT)\Bigr\{2smX\cos(mX)+[(mX)^2(s-2)-2s]\sin(mX)\Bigr\}\biggr].
\end{equation*}
and the imaginary part for the commutator is given by

\begin{equation}\label{ImR}
	\Im[\Delta_{2}]=\frac{\beta}{8\pi m^{s-1} (mX)^{4}}\biggr[\frac{T}{X}\cos(mT)\Bigr\{4mX[(mX)^2(s-1)-6s]\cos(mX)
\end{equation}
\vspace*{-0.6cm}
\begin{equation*}
	\hspace{-0.9cm}+[(mX)^4(s-2)-4(mX)^2(3s-1)+24s]\sin(mX)\Bigr\}+2mX\sin(mT)\Bigr\{2smX\cos(mX)+[(mX)^2(s-2)-2s]\sin(mX)\Bigr\}\biggr].
\end{equation*}
The anti-commutator violates microcausality since it remains finite,
\begin{equation}\label{Re T-0}
	\Re[\Delta_{2}(m,X,T;s)\Big|_{T=0}=-\frac{\beta}{4\pi m^{s-1}(mX)^3}\biggr[\bigr[(mX)^2(s-2)-2s\bigr]\sin(mX)+2smX\cos(mX)\biggr];
\end{equation}
however, the commutator respects microcausality as expected, since it vanishes for arbitrary $m$, $X$ and $s$,
\begin{equation}\label{Im T-0}
	\Im[\Delta_{2}(m,X,T;s)]\Big|_{T=0}=0.
\end{equation}
According to the expressions (\ref{ReR}) and (\ref{ImR}), the only possible critical point for the propagation may be $X=0$; however the  limits are well defined
\begin{equation}\label{Re X-0}
	\lim_{X\rightarrow 0}\Re[\Delta_{2}(m,X,T;s)]=\frac{\beta}{4\pi m^{s-1}}\left[\frac{3s-10}{30}\:mT\sin(mT)+\frac{s-6}{3}\:\cos(mT)\right],
\end{equation}
\begin{equation}\label{Im X-0}
	\lim_{X\rightarrow 0}\Im[\Delta_{2}(m,X,T;s)]=\frac{\beta}{4\pi m^{s-1}}\left[-\frac{3s-10}{30}\:mT\cos(mT)+\frac{s-6}{3}\:\sin(mT)\right];
\end{equation}
these expressions describe the propagation close to the time-like axis; the amplitudes are finite and are increasing as $T\rightarrow+\infty$; for $s=\frac{10}{3}$ one obtains the usual sinusoidal oscillations in this region.

In the figures below the amplitudes (\ref{ReR}) and (\ref{ImR}) are described on the light-cone ($T=X$) for different values of $s$; the case $s=2$ is special in the infrared regimen, due to it corresponds to the only amplitude that drops off for large argument; for $s\neq2$ all the amplitudes remain oscillating without decreasing; moreover, the amplitudes of the oscillations are increasing as $s\rightarrow+\infty$, showing the expected non-flattening. In spite of the non-flattening effect, the intercalation of the amplitudes can be interpreted as a sign of stability for the amplitude with $s=2$, since a small change in the parameter $s$ implies a small change in the amplitudes.

\begin{figure}[h!]
	\centering
	\includegraphics[width=0.45\linewidth]{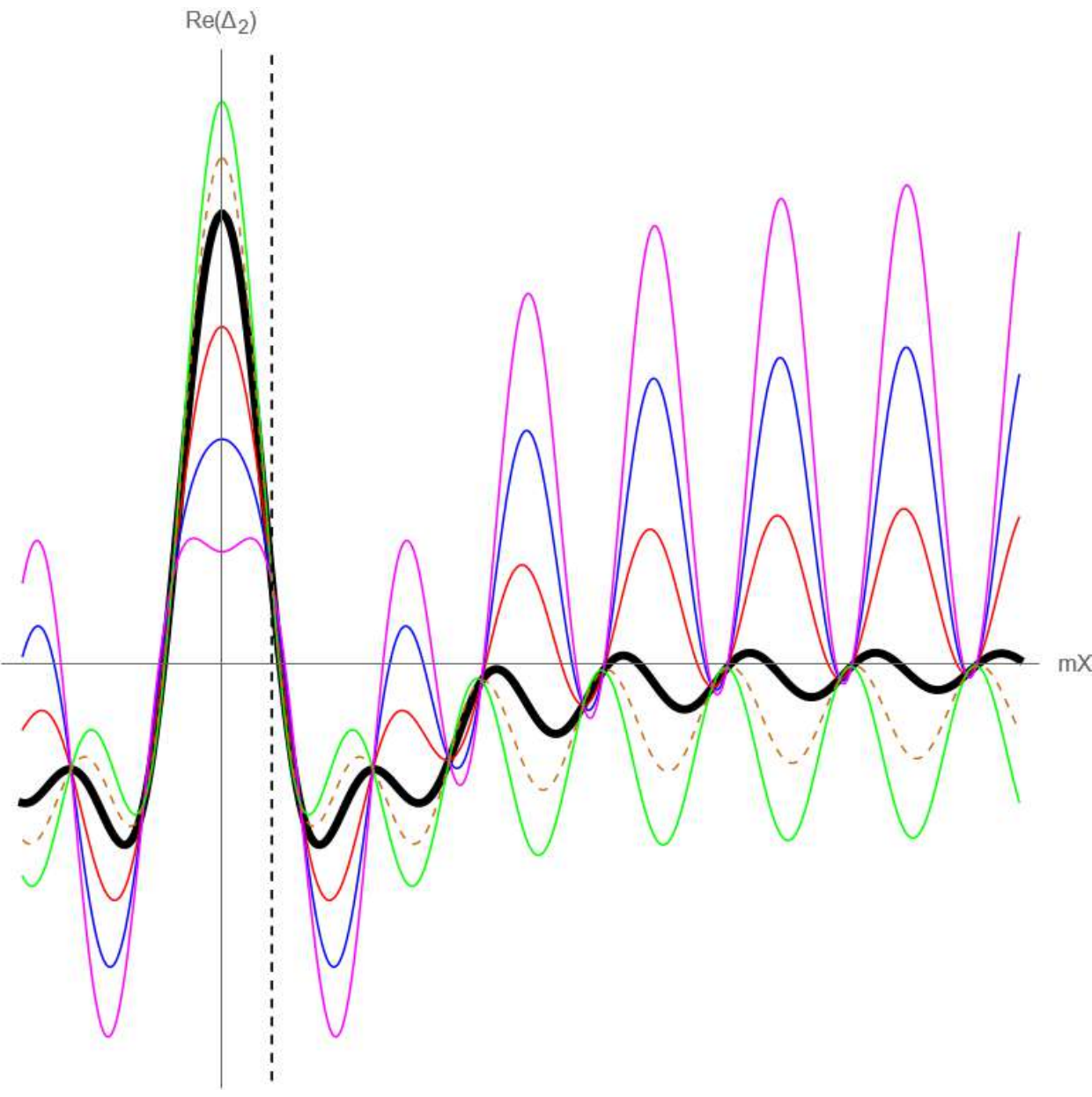}
	\caption[Corto]{The anti-commutator (\ref{ReR}) for fixed $m$ on the light-cone, $T=X$, for $s=1,1.5,2,3,4,5$ with the color code, $1$-green, $2$-black, $3$-red, $4$-blue, $5$-magenta, the dotted curve represents $s=1.5$. Note that all the amplitudes intersect to each other at the same points; in particular 
the first intersection point  occurs at the scale of the Compton wavelength, $X\approx 1.26\frac{1}{m}$, and it is indicated by the dashed vertical line. The first root for $s=2$ occurs also at the Compton wavelength, $X\approx 1.46\frac{1}{m}$.}
	\label{fig:ACDS-11}
\end{figure}
\begin{figure}[h!]
	\centering
	\includegraphics[width=0.5\linewidth]{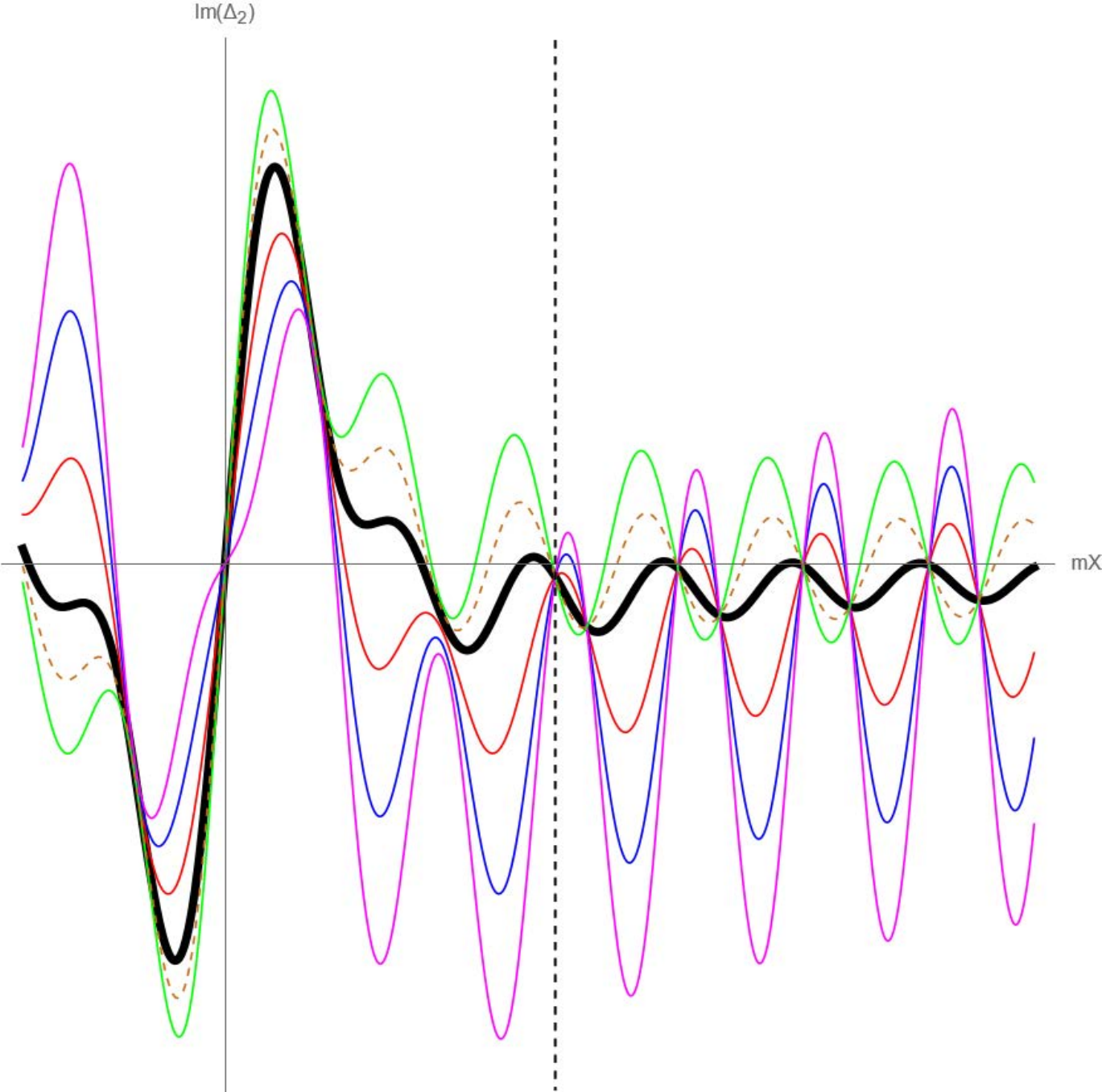}
	\caption{The commutator (\ref{ImR}) for fixed $m$ on the light-cone, $T=X$, for $s=1,1.5,2,3,4,5$ with the color code, $1$-green, $2$-black, $3$-red, $4$-blue, and $5$-magenta, the dotted curve represents $s=1.5$. Note that all the amplitudes intersect to each other at the same points; the first intersection point occurs around eight times the Compton wavelength $X\approx 8.16\frac{1}{m}$, and it is indicated by the dashed vertical line. The first root for $s=2$ occurs at $X=4.86\frac{1}{m}$.}
	\label{fig:CDS-11}
\end{figure}

\newpage
For the commutator described in the figure \ref{fig:CDS-11}, the Feynman propagator for $s=2$ and with $X>0$ is positive definite in the region 
between the origin and the first root (around five times the Compton wavelength); beyond this root, the amplitude oscillates with positive and negative values and drops off for large argument.

The figures \ref{fig:ACDS-11} and \ref{fig:CDS-11} must be compared with the figures \ref{prop less1} and \ref{prop greater1}, for which we recall that the exact amplitudes on the light-cone are well defined for $s>0$, and in particular they are not divergent at $s=2$; instead the $\delta X$-corrections close to the light-cone show a divergence at $s=2$. However the figures \ref{Tr IR} and \ref{Ti IR} below show that the $\delta X$-corrections are well defined for $s=2$ in the infrared limit. Hence, one can consider that the contributions from large momenta, in particular the UV contributions, generate the divergence at $s=2$ for the $\delta X$-corrections in the figures \ref{prop less1} and \ref{prop greater1} in the full description.

Once we have realized that the case $s=2$ is the only one with convergence for large argument, we describe now the corresponding $\delta X$-corrections in the figures \ref{Tr IR} and  \ref{Ti IR}; the corrections turn out to be finite in the full range, and also converge for large argument.

\begin{figure}[h!]
	\centering
	\includegraphics[width=0.8\linewidth]{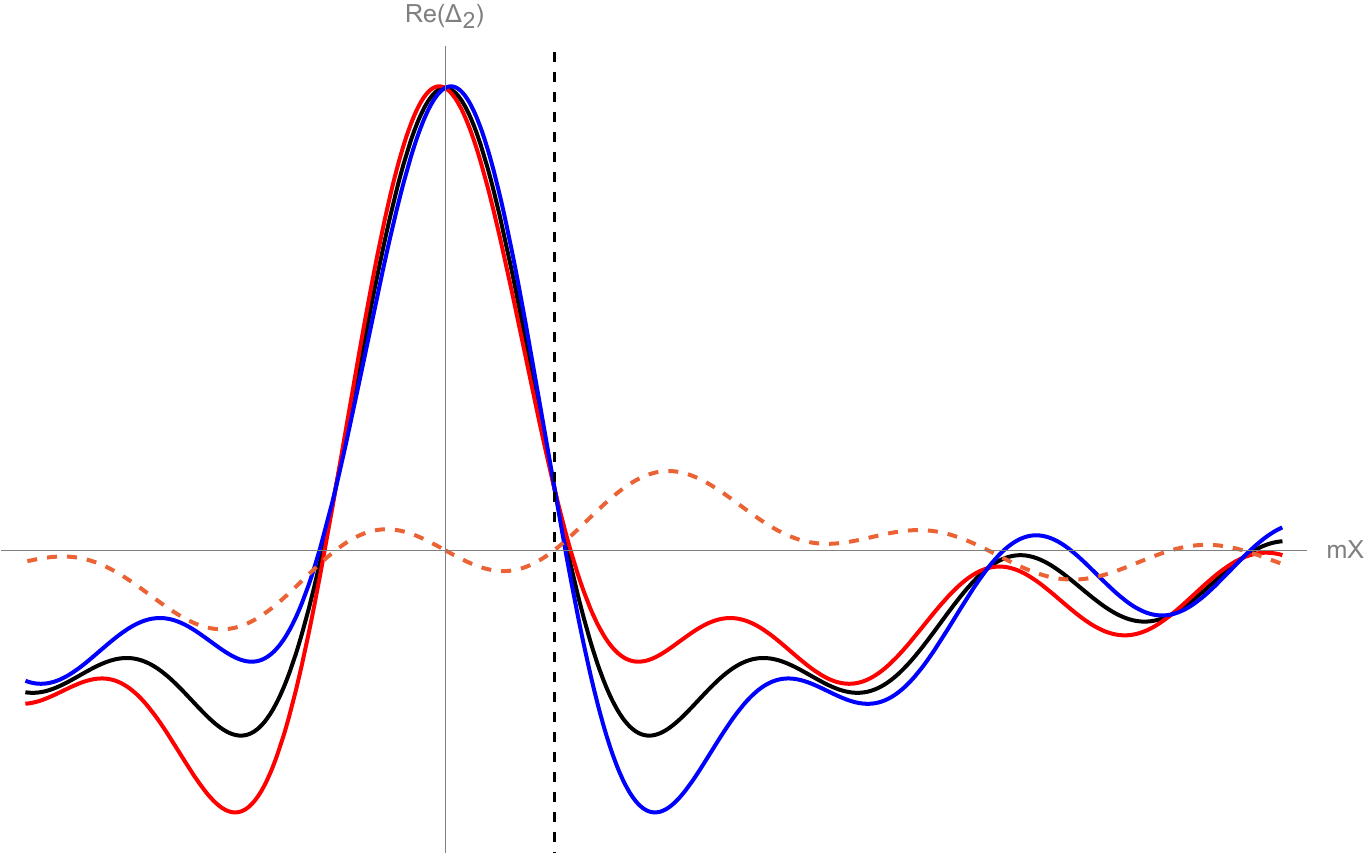}
	\caption{The black curve represents  a zoom of the black curve in the figure \ref{fig:ACDS-11} for the anti-commutator;
	the $\delta X$-corrections are represented by the blue curve (inside the light-cone) and by the red curve (outside the light-cone) with 
 $|\delta X|=0.2$. The dashed line corresponds to the correction term. The first intersection point  is around $X=1.31\frac{1}{m}$; before this point the (classical) hierarchy inside-on-outside the light cone is respected; after the intersection, such a hierarchy is inverted, close to the first roots for the amplitudes. Then the amplitudes take negative values and maintain a competition for large argument.}
	\label{Tr IR}
\end{figure}
\newpage
\begin{figure}[h!]
	\centering
	\includegraphics[width=0.8\linewidth]{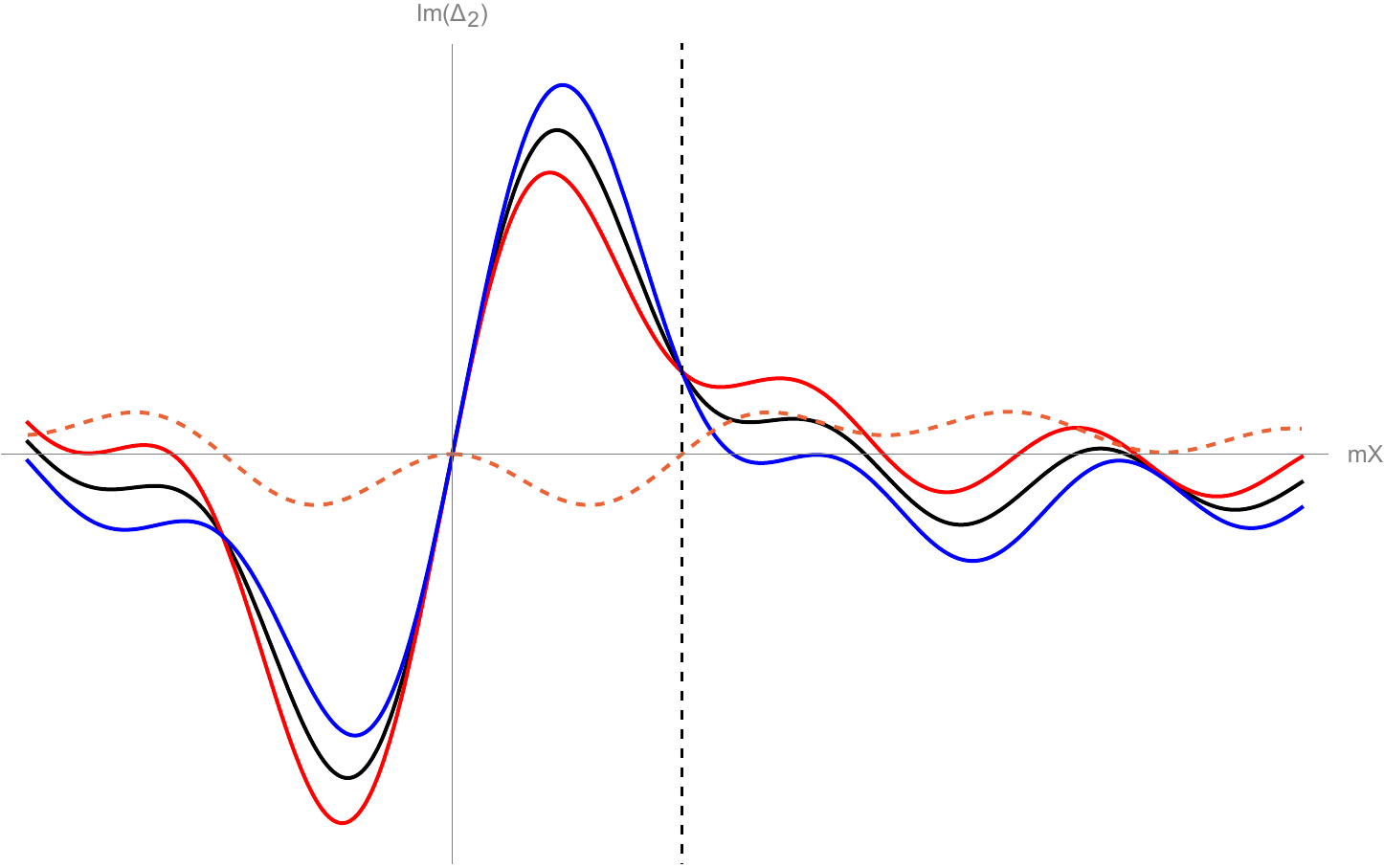}
	\caption{The black curve represents a zoom  of the black curve for the commutator  in the figure (\ref{fig:CDS-11});
	the $\delta X$-corrections are represented by the blue curve (inside the light cone) and by the red curve (outside the light cone) with 
 $|\delta X|=0.2$; the dashed curve corresponds to the correction term. The first intersection point  is around $X=2.70\frac{1}{m}$; before this point the (classical) hierarchy inside-on-outside the light cone is respected; after the intersection, such a hierarchy is inverted. Then the amplitudes take negative values and maintain a competition for large argument.}
	\label{Ti IR}
\end{figure}
Note that all amplitudes described in the figures \ref{Tr IR} and  \ref{Ti IR} are finite in the full range; this finite infrared regime must be compared with 
the divergent $\delta X$-corrections  in the table 2 for $s=2$, for which the divergence can be associated with the large momenta contributions in the full description. Note also that in this infrared regime the stability criterion through the correspondence {\it small deviations near the light cone/ small deformations for the amplitudes} is satisfied; although the classical hierarchy is not respected, this description of amplitudes with close values in the vicinity of the light cone may be physically admissible. 

\section{Concluding remarks}

Preliminary results in three and four dimensions are in order; similar to the $1+1$ dimensional case discussed here, these scenarios also correspond to finite QFT's within a specific range for the parameter $s$. However, there exist important differences, since the three dimensional case requires to invoke the so called holonomic sequences in terms of the Gamma function and the Bessel functions, which show that the amplitudes on the light cone are positive definite in the range  $2\leq s\leq 4$ (which corresponds to the universal restriction $2\leq s<d+1$), and they drop off for large separation between the two correlated points in the propagation. In contrast, the four dimensional case is curiously simpler than the three dimensional case, since it does not require holonomic sequences, and it is technically closer to the case $1+1$ developed here.
These cases will be reported in detail elsewhere.

Once the Lorentz symmetry is broken with a nontrivial weighted measure, it remains to identify the (quantum) residual symmetry; the example of the doubly special relativity with a deformed dispersion relation, and with a nonlinear version for the Lorentz symmetry   \cite{smolin}, may serve as a guide in this exploration. In particular for the case $\frac{M}{m}<1$, the limit $s\rightarrow+\infty$  leads to vanishing weighted amplitudes, and then to a trivial QFT without quantum propagation; this limit with a maximally broken Lorentz symmetry, corresponds to a sort of classical limit in the $s$-space, which is different to the standard limit with $\hbar \rightarrow 0$;
the identification of the corresponding residual symmetry is mandatory, and it will be considered in forthcoming works.

In this work we have considered only the free propagator for a massive scalar particle; interacting field theories for higher spin will be considered in the future. Similarly the impact at perturbative level of our re-formulation of the propagator field theory will be explored in forthcoming communications.

{\bf Acknowledgements:}
This work was supported by the Sistema Nacional de Investigadores (M\'exico).  S.G.S and R.B.R would like to thank VIEP-BUAP (M\'exico) for financial support. Computations have been made using Mathematica.

\end{document}